\documentclass[aps,preprint,showpacs]{revtex4}  
\usepackage{graphicx,epsfig}  
\usepackage{dcolumn}      
\usepackage{amsmath}        
\usepackage{bm}

\begin{document}      
             
\preprint{Editorially approved for publication at Phys. Rev. B}     
        
\title{Controlling many-body effects in the midinfrared gain and THz       
absorption of quantum cascade laser structures}        
        
\author{M.F. Pereira Jr.}         
\email{mauro.pereira@nmrc.ie}
      
\affiliation{NMRC, University College, Lee Maltings, Prospect Row, Cork,    
Ireland }      
\author{S.-C. Lee and A. Wacker$^{\dagger}$ }
\affiliation{Institut f{\"u}r Theoretische Physik,             
Technische Universit{\"a}t Berlin,  Hardenbergstr. 36,       
10623 Berlin, Germany.} 
        
\date{\today}      
        
\begin{abstract}       
A many-body theory based on nonequilibrium Green functions, 
in which transport and   
optics are treated on a microscopic quantum mechanical   
basis, is used to compute gain and absorption in the optical and THz regimes 
in quantum cascade laser structures. The relative importance of Coulomb   
interactions for different intersubband transitions depends strongly on the   
spatial overlap of the wavefunctions and the specific nonequilibrium   
populations within the subbands. The magnitude of the Coulomb effects can be
controlled by changing the operation bias.   
       
\end{abstract}        
        
        
\maketitle   

\section{Introduction}
The quantum cascade laser (QCL) has evolved, since its first  experimental   
realization \cite{FAI94a}, into an important device for infrared   
applications.   
Recently, lasing   
in the THz regime has been demonstrated, thus opening further    
perspectives \cite{KOE02}.  
The operation of these semiconductor heterostructure devices     
is based on  optical intersubband transitions   
within the conduction band, which are the focus of this work.   
The importance of many-body effects on interband transitions is a well   
established fact \cite{Zim}. 
The intersubband case has also been  
studied both experimentally and 
theoretically  \cite{Helm:99} and is a topic of current research \cite{Li:03,Ines:03}.
Typically, idealised cases of isolated quantum wells  with equilibrium   
distributions for the carriers are considered in a two  subband scenario, to   
facilitate the computational challenges. Assuming that only one   
level is occupied, exchange and depolarisation effects were found to be very   
relevant for both gain and absorption \cite{NIKO97}. Recent work has   
demonstrated the importance of fully solving the integral equation for the   
optical susceptibility for quantum wells 
\cite{Li:03,Ines:03,Ning:02,Gumbs:95,Faleev:02}.  
While standard semiconductor lasers based on interband transitions can be   
modelled by equilibrium distributions for electrons and holes with different   
quasi-Fermi levels \cite{PER98}, this is not appropriate for QCLs   
where a full kinetic theory is necessary to correctly determine the    
nonequilibrium populations of the multiple level system.    
     
The main issues of our paper are: (i) the question how these many-body effects
are affected by the inherent nonequilibrium distribution in a QCL, where 
the  interplay of a large number of different levels is crucial for the 
operation of the device; 
(ii) to what extent many-particle corrections 
can be modified in a given structure out of equilibrium and in a complex
wavefunction scenario.

We show that a characteristic QCL 
design provides rather small overall Coulomb corrections to the gain 
transitions in the infrared, while significant level shifts occur in the 
THz absorption region. The latter depend strongly on the bias applied to the 
structure, which   
opens a possibility for a systematic variation of Coulomb effects in an   
experimental study of a complex nonequilibrium system. 
Here a wider variation of parameters is possible compared with 
previous experimental studies in quantum wells by either
applying a bias \cite{Tsu:00,Luin:01} or optically pumping a laser
\cite{Liu:00}.
  
Our findings thus give evidence of the need of our more advanced approach 
to model complex nonequilibrium studies. The study of simple 
structures in ref.~\cite{NIKO97} would suggest similar strengths of 
Coulomb corrections
in both gain and absorption regimes. Although those results are
qualitatively correct for the cases previously studied, they can
not be used for more complex structures of current interest, as demonstrated
in our analysis, since the Coulomb matrix elements and occupation functions
are extremely different for gain and absorption transitions. That large
discrepancy is explained here for the first time to the best of our knowledge.

\section{Main Equations}  
Our calculations are based on the stationary state under   
operating conditions which is obtained by a self-consistent  
solution of the quantum  kinetic equations \cite{LEE02a}.  
Here scattering with phonons and impurities   
are included via the respective self-energies in the Dyson equations, and   
electron-electron interaction is treated in mean field approximation. The   
lattice temperature only enters via the phonon bath at 77 K, which is assumed   
to be in thermal equilibrium, neglecting hot phonon effects   
\cite{COM02}. For each subband, the nonequilibrium occupation   
functions, $f_{\nu}(k)$ are calculated from the diagonal terms in the carrier  
Green functions $G^{<}_{\nu \nu}({\bf k},\omega)$ which are expanded in terms   
of eigenstates $\phi_{\nu}(z)$ of the single particle Hamiltonian including
the mean field. (These states will be referred to as levels in the following.)
The calculated dephasing $2 \Gamma_{\nu}$ is the FWHM   
of the spectral function, ${\Im}\{G^{\rm ret}_{\nu\nu}({\bf k=0},\omega)\}$   
and $E_{\nu}({\bf 0})$ the corresponding location of the peak of the spectral   
function so that the electronic states are well defined quasi-particles, whose   
properties are evaluated on a microscopic basis.      
The optical absorption $\alpha$ at a given photon energy $\hbar \omega$ can be   
calculated from the imaginary part of the optical susceptibility,   
\begin{eqnarray}      
\alpha(\omega) = \frac{4 \pi \omega}{c n_{b}} {\Im}\{\chi(\omega)\}, \; \;   
\chi(\omega) = 2 \sum_{\mu \ne \nu,\vec{k}} \, \wp_{\mu \nu} \,   
\chi_{\nu,\mu}(k,\omega).   
\label{EqAbsorb}     
\end{eqnarray}      
Here $n_{b}$ denotes the background refractive index, $c$ is the speed of   
light, $\wp_{\nu \mu}$ is the transition dipole moment between the subbands   
$\nu$ and $\mu$, which are labelled $\nu=1,2,\ldots $ with increasing   
energy. In our calculation we select the 9 lowest subbands per period. The   
definition of a period (and thus the choice of level 1) is of course   
arbitrary in such structures and we must also take into account transitions   
between neighbouring periods.  Numbering the levels in the next period to   
the left by 10--18, we consider all transitions $(1,\ldots 9)\leftrightarrow   
(1,\ldots 18)$  in the computations.  We evaluate the susceptibility   
function, $\chi_{\nu \mu}(k,\omega)$, by the carriers' Keldysh Green's function,   
$G$, whose time evolution is described by a Dyson equation, with Coulomb   
interactions as well as other scattering  mechanisms included in self-energies,   
$\Sigma$ \cite{PER98}. The general  theory includes carrier-carrier   
scattering, diagonal and non-diagonal  dephasing. In the numerical results   
presented here, we keep only the subband shift,
depolarisation and exchange terms in the   
electron-electron interaction  (Hartree-Fock). 
Similarly to the interband case, \cite{PER98} but with additional terms due to   
depolarisation as well as full consideration of a Hartree (mean field) term in   
the quasi-free particle term $E_{\nu}(\bf k)$, the resulting integral equation   
for the susceptibility function reads,   
\begin{widetext}   
\begin{multline}   
\left[\hbar  \omega - e_{\nu \mu}({\bf k}) + i \left(   
\Gamma_{\mu}+\Gamma_{\nu} \right) \right]  \chi_{\nu\mu}({\bf k},\omega)+   
\left( f_{\nu}({\bf k})-f_{\mu}({\bf k}) \right) \; 2 \; V \left( { \nu \; \mu   
\; \mu \; \nu \atop 0 } \right) \; \sum_{{\bf k}'} \chi_{\nu\mu}({\bf   
k}',\omega) \\  - \left( f_{\nu}({\bf k})-f_{\mu}({\bf k}) \right) \;   
\sum_{{\bf k}'\ne {\bf k}} \chi_{\nu\mu}({\bf k}', \omega) \;  V \left( { \nu   
\; \nu \; \mu \;  \mu \atop {\bf k} -{\bf k}' } \right) = \wp_{\nu \mu} \left(   
f_{\nu}({\bf k})-f_{\mu}({\bf k}) \right) \, .   
\label{EqHarFock}      
\end{multline}   
Terms with more then two different indices vanish identically in the limit of   
isolated wells with infinite confinement barriers and remain small in the more   
complex structure studied here.  The bare Coulomb interaction is given by   
\begin{equation}
\label{Matrixelements}      
V \left( { \mu \; \nu \; \lambda \; \beta \atop {\bf k}-{\bf k}' } \right) =   
\int dz \, dz'\, \phi^{\ast}_{\mu} (z)  \phi_{\nu} (z)  \frac{2 \pi e^2\,   
\exp{\left(-|{\bf k}-{\bf k}' | | z - z' | \right) } } {\epsilon_0 A|{\bf k}-{\bf k}' |} \,   
\phi^{\ast}_{\lambda} (z') \phi_{\beta} (z'),   
\end{equation}      
where we have introduced the normalisation area $A$, and the background   
dielectric constant $\epsilon_{0}$. 
The explicit expressions actually programmed here are given in the Appendix. 
\begin{equation}\begin{split}      
e_{\nu \mu}({\bf k}) =&E_{\nu}({\bf k}) -E_{\mu}({\bf k}) - \sum_{{\bf   
k}'} f_{\nu}({\bf k}')  \; V \left( { \nu \; \nu \; \nu \; \nu \atop {\bf   
k}-{\bf k}' } \right)  +\sum_{{\bf k}'} f_{\mu}({\bf k}')  \; V \left( { \mu   
\; \mu \; \mu \; \mu \atop {\bf k}-{\bf k}' } \right) \\ &+ \sum_{{\bf k}'}   
[f_{\nu}({\bf k}')-f_{\mu}({\bf k}')]  \; V \left( { \nu \; \mu \; \mu \; \nu   
\atop {\bf k}-{\bf k}'} \right)\, ,   
\label{Energies}      
\end{split}\end{equation}     
\end{widetext}   
is the energy difference between the levels renormalised   
by the exchange interaction, which we refer to as subband shift in the   
following.  The second term on the left-hand side of Eq.~(\ref{EqHarFock})   
gives rise to the depolarisation shift, while the last term (exchange   
contribution) is analogous to the excitonic coupling term in interband   
transitions \cite{PER98}. The equations above reduce to those of     
Ref. \cite{CHU92} in the two-band limit at equilibrium.   
Note that by running over all subband indices $\nu \ne \mu$ in   
Eqs.~(\ref{EqAbsorb},\ref{EqHarFock}), we do not apply the rotating wave   
approximation.   

\section{Numerical Results and Discussion}

\begin{figure}   
\includegraphics[width=0.95\columnwidth]{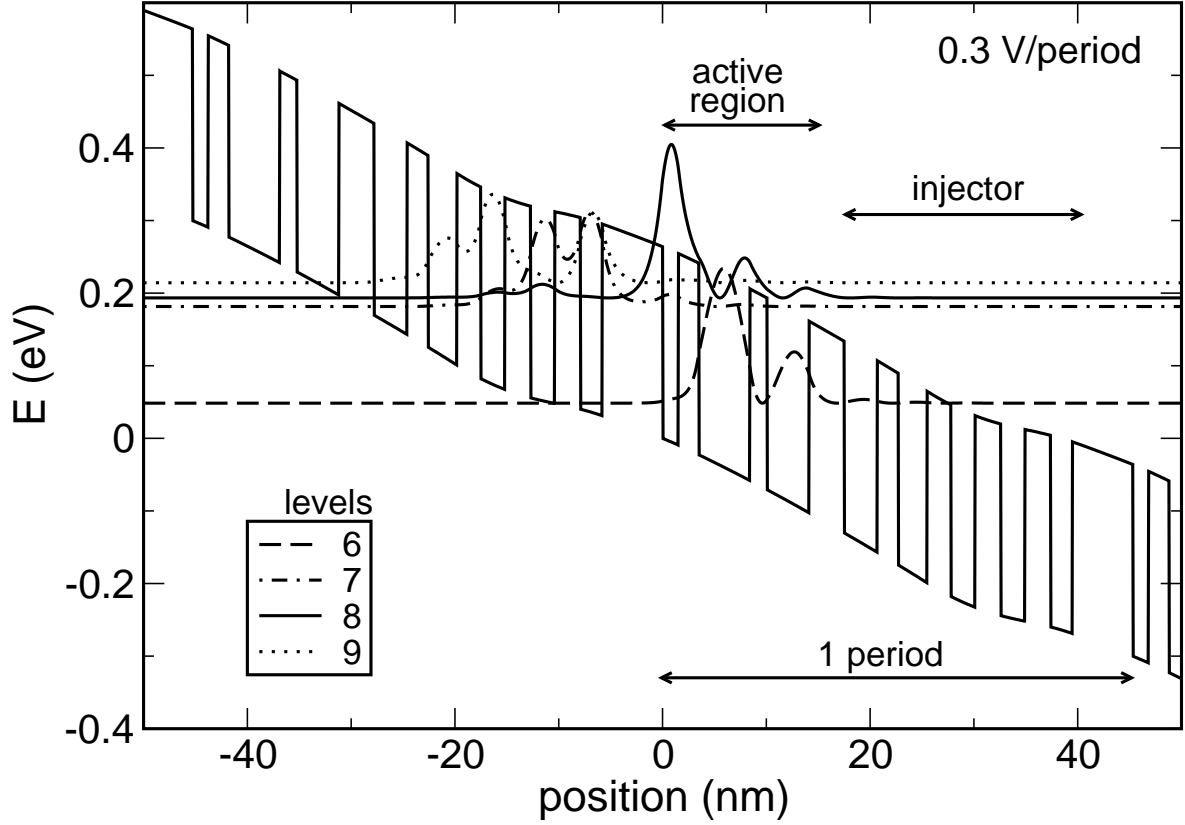}

\caption{Section of the quantum cascade laser structure and a few    
relevant levels within a period, including mean field
effects due to the ionised dopants in the structure, and for a bias V=0.3 V/period.
Level 7 is an injector level. Levels   
6 and 8 yield the main contribution for the midinfrared gain 
spectrum, while   
levels 7 and 9 are responsible for the main THz absorption feature.
}   
\label{fig1}     
\end{figure}

For numerical applications of the theory, we choose the very successful   
prototype introduced by Sirtori {\em et al.} \cite{SIR98}, see   
Fig.~\ref{fig1}. From a systematic study of doping   
dependence \cite{GIE03} it was found that the performance is best for a   
doping around  $6.4\times 10^{11}/\textrm{cm}^2$ which is used throughout 
this work together with a lattice temperature of 77 K unless stated otherwise.

\begin{figure}     
\includegraphics[width=0.95\columnwidth]{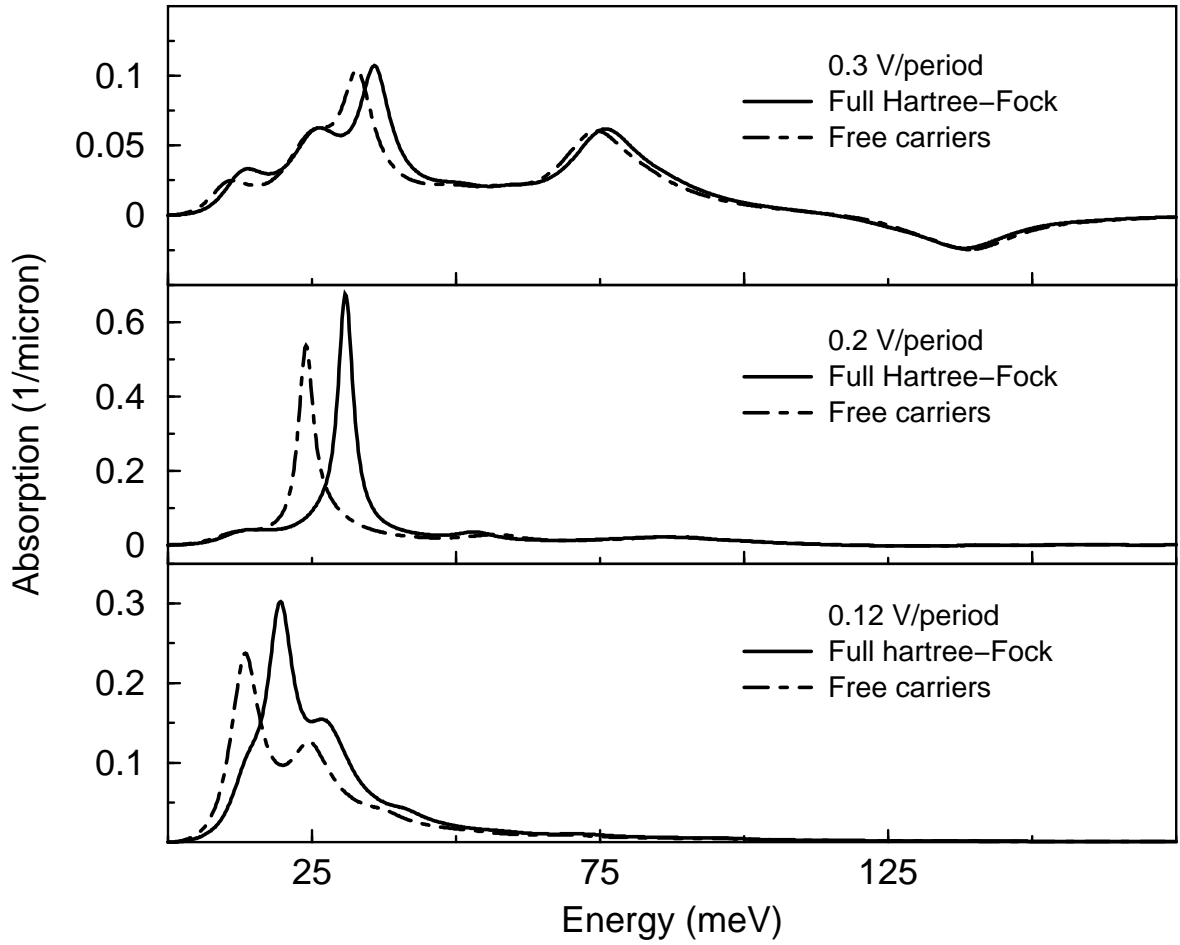}
\caption{Influence of bias  on absorption/gain spectra. From top to bottom the bias is V=0.3, 0.2 and 0.12 V/period. In all panels, the dot-dashed lines
are for free carriers, while the solid curves have full Hartree-Fock Coulomb 
corrections.}

 \label{fig2}     
\end{figure}

In Fig.~\ref{fig2} we compare the absorption spectra  evaluated  
with and without Coulomb effects for different biases. We find that  gain sets in around 0.2 V/period at energies $\hbar \omega\approx 130$ meV   
(infrared region) in good agreement with the experimental findings   
\cite{GIE03}. In addition there is a pronounced absorption around  30 meV   
(THz-region).  While the Coulomb effects are hardly visible in the gain   
region, significant shifts of peak height and position occur in the THz   
absorption lines, in particular at low bias.  
    
In the following we present a systematic study of the Coulomb terms in
Eq.~(\ref{EqHarFock}), which crucially depend on the Coulomb matrix elements
and the occupation differences, which are displayed in Table \ref{tab2}
for the dominant transitions.  A glance at the  values $n_{\nu}-n_{\mu}$
[which roughly scale with $f_{\nu}({\bf k})-f_{\mu}({\bf  k})$] shows that the
structure is out of equilibrium,  as necessary for laser operation. In order
to quantify changes in Coulomb effects, we express the matrix element for the
depolarisation term as 
$L^{dep}_{\nu \mu} = \frac{ 9 \pi \epsilon_{0} A}{20 e^{2}} \, 
V^{\nu \mu \mu  \nu}_{0}$, so that it reduces to the well width in
the limiting case  of a quantum well with infinite confinement barriers for
transitions  between the two lowest subbands. This reflects the common
knowledge that the depolarisation shift increases with well width,
i.e. increasing spatial extension of the  wave    functions.  For a QCL the
scenario is much more involved, as the spatial    overlap of the wave
functions essentially contributes to the Coulomb effects    as well. Similar
quantities can also be defined for the exchange and subband shift,  namely
$L^{ex}_{\nu \mu}$, and $L^{bg}_{\nu \nu}$. Explicit expressions are derived
in the Appendix and numerical values are given in Table~\ref{tab2}.  In
contrast to $L^{dep}$, both $L^{ex}$, and $L^{bg}$ are inversely proportional
to the corresponding Coulomb matrix elements.  We can then expect that
relatively  large $L^{dep}$ and small $L^{ex}$ should correspond to large
dipole moments,  or in other words, to strong overlap between the
wavefunctions of the levels involved in the transition. Only the wavefunction
for  level $\nu$ appears in $L^{bg}_{\nu \nu}$. Thus, its value remains within
the same order of magnitude irrespective of the transition.  Note that the
significant difference between the values of $L^{dep}, L^{ex}$, and $L^{bg}$
as well as for the different transitions further demonstrates that results
from a simplified two-level system in a quantum well, where these lengths are
identical by definition, cannot be easily generalised to more complex
scenarios, like the real QCL structure discussed here.

In the THz region, $L^{dep}_{\nu \mu}$ and the dipole moment $\wp_{\nu \mu}$
tend to decrease with increasing bias, due to the changes in the wave
function overlap.  Furthermore the population decreases in the injector. This
explains  the general trend that the Coulomb effects become weaker with
increasing bias.  At 0.2 V/period, the concentration of the population in the
lowest injector  level 7 leads to a very strong absorption feature in the THz
region although  $L^{dep}_{\nu \mu}$ is smaller than at 0.12 V/period.  In the
infrared region,  the larger bias increases the occupation of the subbands
which are  responsible for laser action. Furthermore both $L^{dep}_{\nu \mu}$
and  $\wp_{\nu \mu}$ increase with bias as the injector levels leak into the
active  region. Nevertheless $L^{dep}_{\nu \mu}$ is much smaller than in the THz
absorption region, and furthermore the compensation between the different
many-particle effects (see below) in the dominating gain transitions leads  to
very  small resulting shifts in Fig.~\ref{fig2}.  The contrast between the
interesting Coulomb shifts in the low energy side and the almost vanishing
shift in the high energy gain side is dramatic. This is the reason why gain
calculations without Coulomb corrections \cite{LEE02a} could explain
experimental gain spectra in this structure reasonably well, but will fail in
the absorption  region.  The main contributions to the absorption in the THz
region are in fact transitions involving the two lowest injector levels at
each bias.  In other words, levels 6 and 7 for 0.12 V/period, levels 7 and 8
for  0.2 V/period and levels 7 and 9 for 0.3 V/period.

\begin{figure}      
\includegraphics[width=0.95\columnwidth]{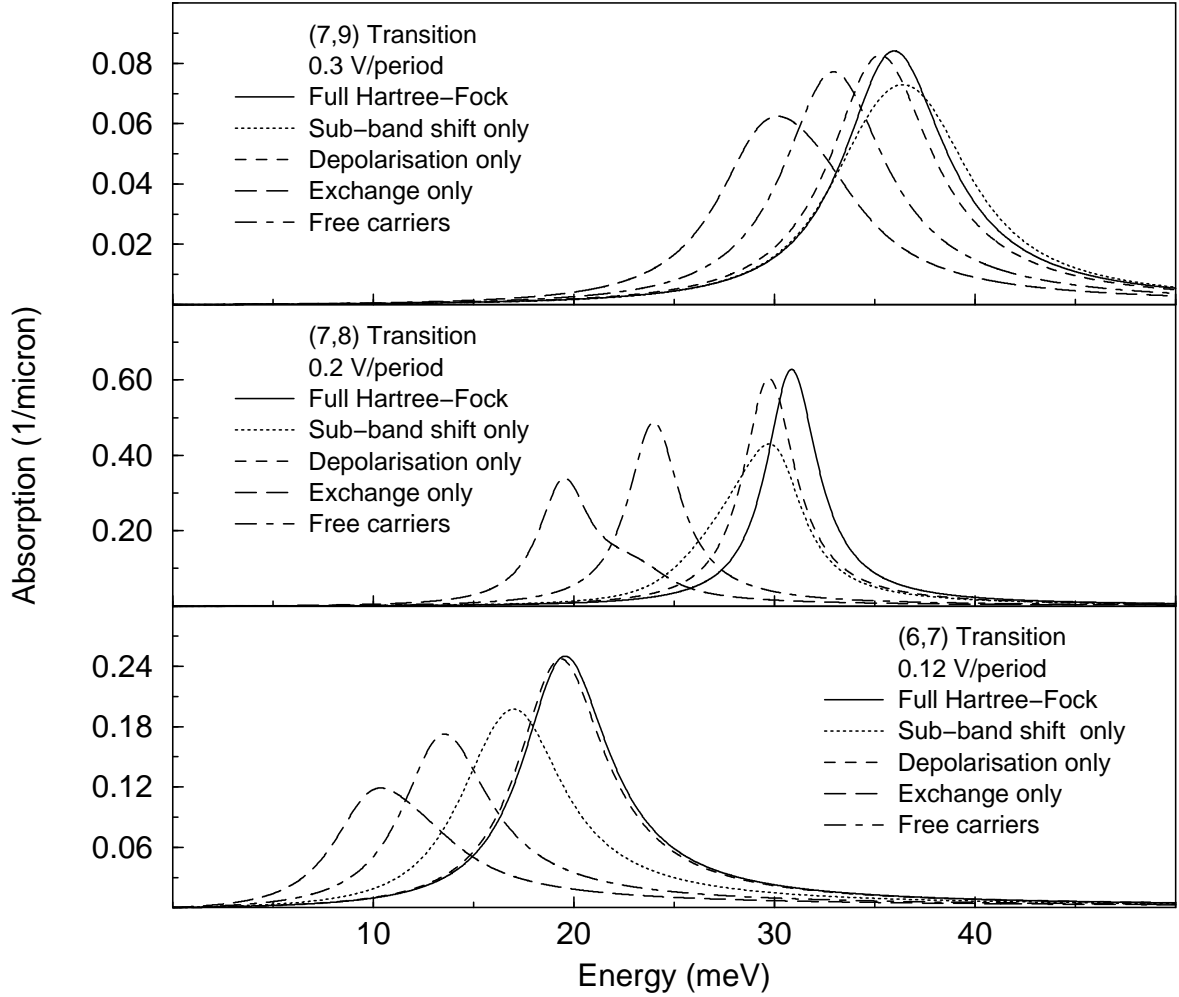}

\caption{Details of THz absorption including mean field effects in the dominant
transition of each bias.  From bottom to top, the bias is V= 0.12, 0.2 and 0.3 V/period, and the corresponding transitions are between levels (6,7), (7,8) and
(7,9) as discussed in the text. In all panels, the solid, dotted, short-dashed, long-dashed and dot-dashed curves are respectively for Full Hartree-Fock, sub-band shift only, depolarisation only, exchange only, and free carriers.}   
\label{fig3}      
\end{figure}

The interplay between the different Coulomb corrections in the THz region is   
analysed in more detail in Fig.~\ref{fig3}.  We select the dominant   
transitions for each bias  to highlight the different contributions more   
clearly. The general tendency is that the exchange contribution imparts a   
red-shift while subband shift and depolarisation terms lead to   
blue-shifts.  The depolarisation shift decreases with bias (in accordance   
with the discussion above), while the tendency of the exchange contribution   
and the subband shift is less clear. These two terms almost cancel each   
other, so that the blue-shift due to depolarisation dominates the behaviour.   
Because the subband shift slightly overcompensates the exchange contribution,
the  full Hartree-Fock curve exhibits a small additional  blue-shift, in
particular for 0.2 V/period. 
   
\begin{figure}      
\includegraphics[width=0.95\columnwidth]{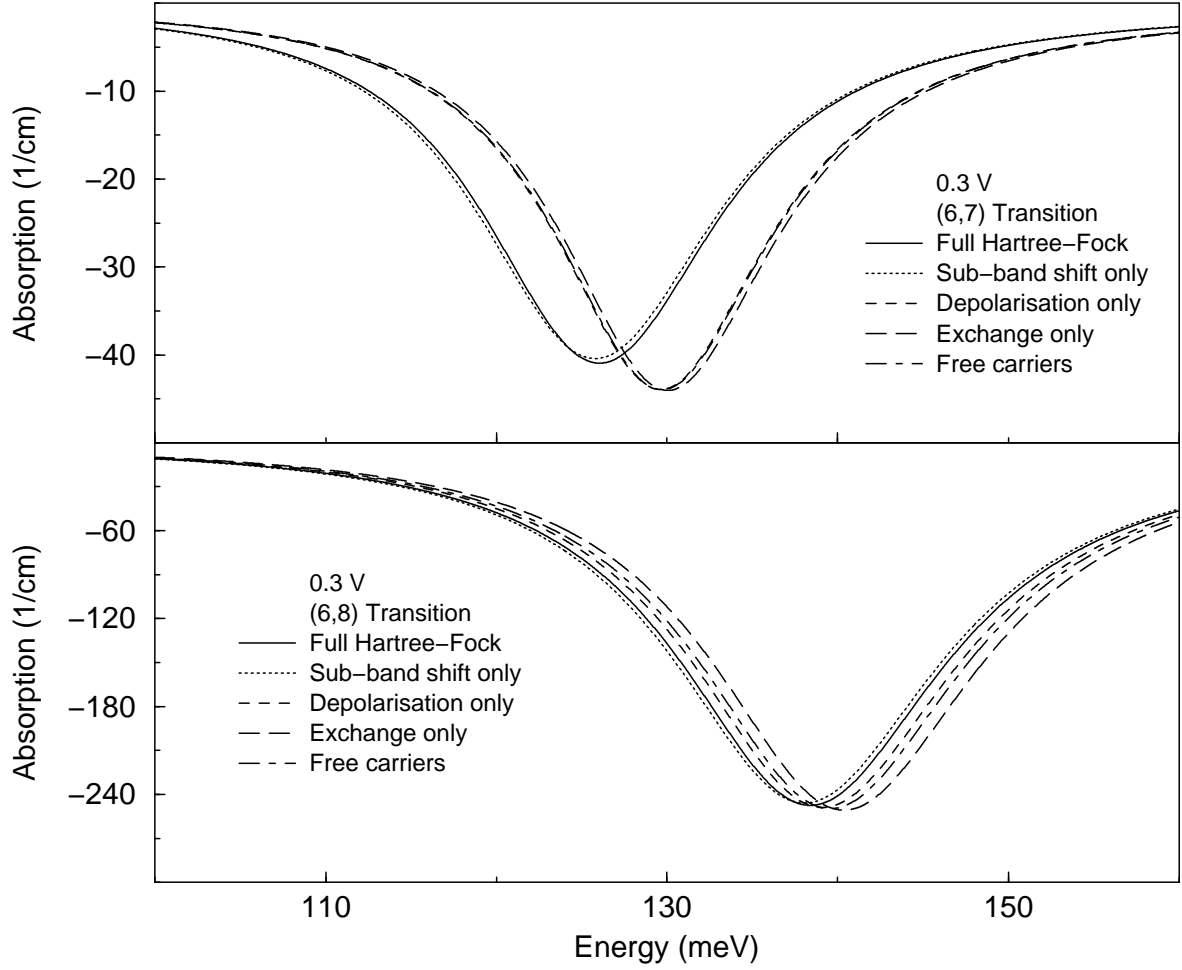}

\caption{Details of the main transitions that contribute to the gain spectra 
including mean field effects for a 
bias V= 0.3 V/period. The lower and top panels are respectively for
(6,8) and (6,7) transitions. In both panels, the solid, dotted, shot-dashed, long-dashed and dot-dashed curves are respectively for Full Hartree-Fock, sub-band shift only, depolarisation only, exchange only, and free carriers.}   
\label{fig4}      
\end{figure}

Figure~\ref{fig4} shows details for the two dominating  gain transitions    at
0.3 V/period.  We notice first that all many-particle effects reverse  sign
for gain transitions. This is due to the sign change in the occupation
differences entering Eqs.~(\ref{EqHarFock},\ref{Energies}) \cite{NIKO97}.
Both transitions exhibit a  red-shift due to all the Hartree-Fock
corrections.  The shift is  stronger for the (6,7) transition although the
depolarisation length is much smaller compared to the (6,8) transition, see
Table~\ref{tab2}. For gain transitions a  significant part of the upper
laser level is located in the injector,  resulting in a weaker spatial
overlap of the wave functions $\phi_{\mu}$ and $\phi_{\nu}$. Thus the
Coulomb matrix elements entering exchange and depolarisation corrections
(such as $L^{dep}_{\nu \mu}$) are reduced compared to the THz absorption
within the injector region.  In contrast Coulomb matrix  elements of the type
$V(\mu,\mu,\mu,\mu)$ enter the subband shift (\ref{Energies}), which
therefore dominates the behaviour in the gain region.  Indeed, a lookup at
Table~\ref{tab2} shows that $L^{dep}$ is small, $L^{ex}$ is large,
consistently with the small dipole transition moment, while $L^{bg}$ is
relatively small, leading to the scenario depicted in Fig.~\ref{fig4}.

The spectra of typical QCLs are found to result from a delicate interplay of
different transitions  with strongly varying transition parameters.  Standard
approaches for idealised quantum wells cannot describe this situation, where
the full nonequilibrium distribution obtained from quantum transport
\cite{LEE02a}, or possibly semi-classical Monte-Carlo \cite{IOT01a}
simulations,  must be combined with the correct wavefunctions for a consistent
description and understanding of  the detailed spectra. Table \ref{tab2}
shows that $L^{dep}_{\nu \mu}$ and $\wp_{\nu \mu}$ can be easily modified
within the same structure by changing the bias. This allows for a
systematic experimental study of the magnitude of Coulomb effects as a
function of transition parameters. Thus  QCLs constitute promising structures
to investigate the influence of Coulomb effects on intersubband transitions. A
central issue of such experimental studies as well as future theoretical
work will be the importance of higher order Coulomb corrections together with
a more stringent treatment of screening compared to our Hartree-Fock
approach. Recent work indicates that compensations between diagonal and
nondiagonal terms \cite{Ines:03} may strongly reduce the importance of
higher-order
electron-electron scattering terms in simple model structures. 
This indicates that neglecting scattering
in our Hartree-Fock approach may be a reasonable approximation,
although this requires further detailed investigation.

Furthermore we illustrate the importance of many body effects
as compared to the free carrier model as a function of both doping densities
and temperatures. In order to keep our discussion as close as possible to
experimental realizations, we chose densities and temperatures compatible with
the samples described in Refs.~\cite{SIR98,GIE03}.

\begin{figure}      
\includegraphics[width=0.95\columnwidth]{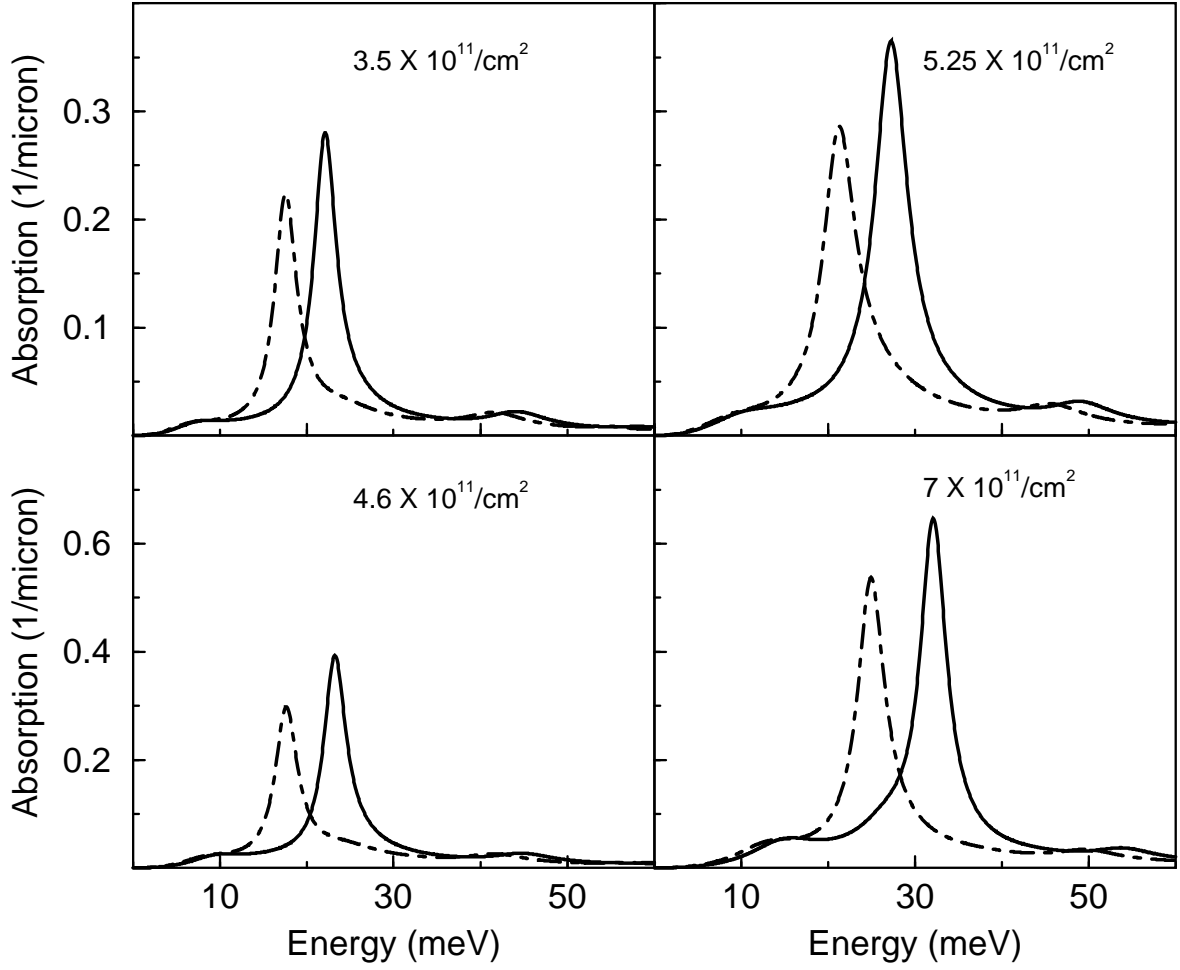}

\caption{Comparison of absorption spectra with (solid) and without 
(dot-dashed) Hartree-Fock Coulomb corrections. In all panels the
bias is V=0.2 V/period. The carrier densities are given
respectively by N=3.5, 4.6, 5.25, and 7 $\times$ $10^{11}$/cm$^{2}$.}
\label{figdens}      
\end{figure}

Figure \ref{figdens}  shows the absorption spectra with and without
Hartree-Fock Coulomb corrections for different doping densities.
Our free
carrier model includes the mean field of the ionised dopants and as the
density increases, they shift the peak of the main absorption transition to
the blue. The many-body effects impart a further
intersubband blue shift and increase in oscillator strength of the main
transitions. 
Although both the subband shift and overall
oscillator strength increase with carrier density, the impact of the Coulomb
corrections is similar for all densities and follows our detailed
analysis of the  N=6.4  $\times$ $10^{11}$/cm$^{2}$ case given above.

\begin{figure}      
\includegraphics[width=0.95\columnwidth]{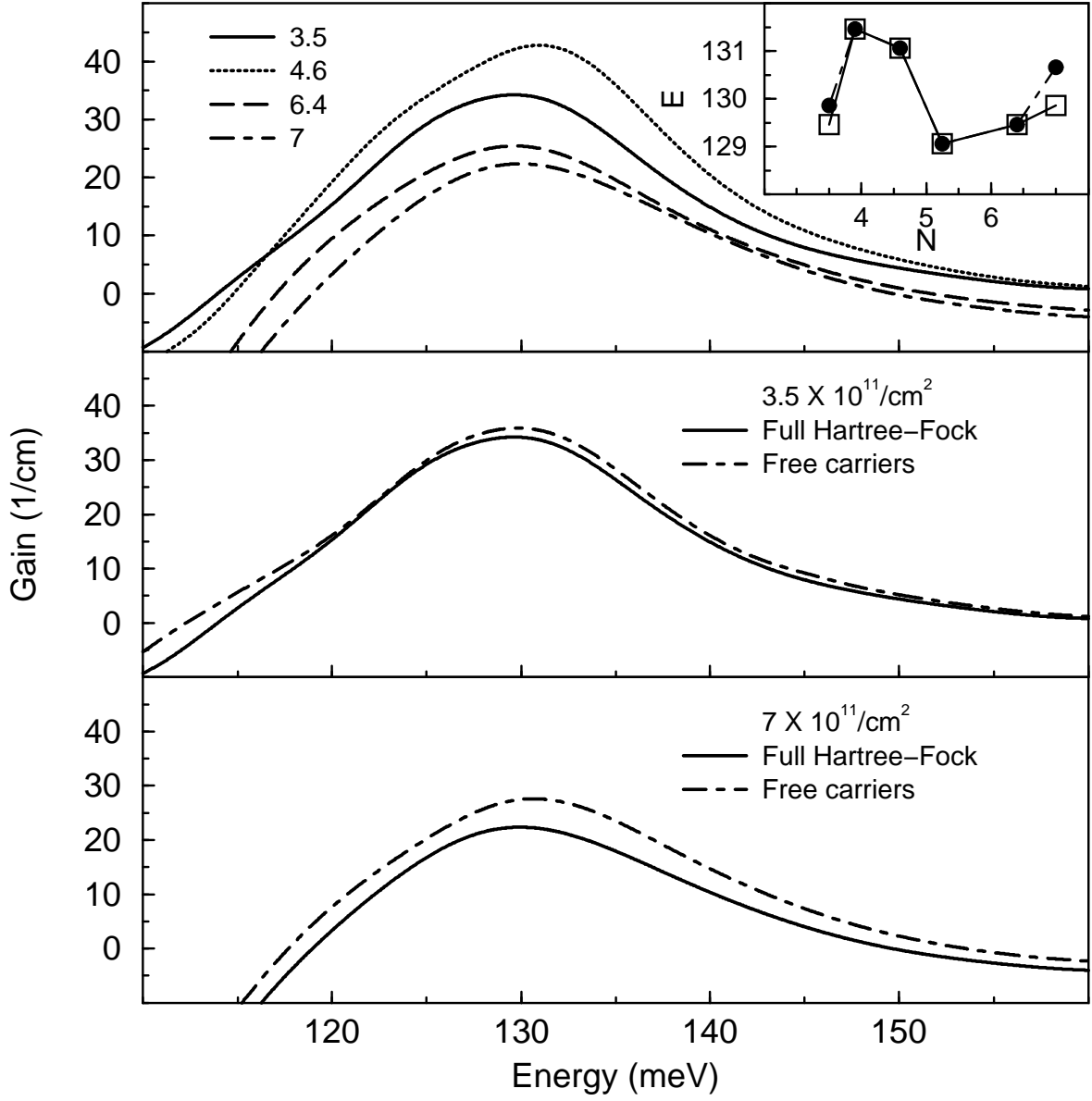}

\caption{Gain spectra as a function of doping. All curves in the top panel
have full Hartree-Fock corrections. The solid, dotted, long-dashed and
dot-dashed curves correspond, respectively to N=3.5, 4.6, 6.4 and 7 $\times$
$10^{11}$ /cm$^{2}$. The inset depicts the spectral position of peak gain (in meV) as a function of doping in $10^{11}$ /cm$^{2}$. The open squares are for
full Hartree-Fock while the circles are for free carriers. The symbols are
connected by solid and dot-dashed lines respectively to guide the eye.
In the central and bottom panels, the dopings are, respectively,
N=3.5 and 7 $\times$ $10^{11}$ /cm$^{2}$. The solid and dot-dashed curves are 
respectively for full Hartree-Fock and free-carrier calculations.}
\label{figgain}      
\end{figure}

The density dependence study is complemented with Fig.~\ref{figgain} that depicts the evolution
of gain (-absorption) spectra with density. Maximum gain is reached around
N=4.6  $\times$ $10^{11}$/cm$^{2}$. The lower panels show that Coulomb corrections are larger for the larger density and the inset depicts the evolution of
peak gain position with doping. Although the corrections are small it is interesting to note that Coulomb corrections lead to smaller increase in peak gain
position on the high density, making the results slightly closer, 
qualitatively, to the experimental findings of Ref.~\cite{GIE03}. However, Coulomb effects only are not enough to explain their results and a more detailed analysis is required.

\begin{figure}      
\includegraphics[width=0.95\columnwidth]{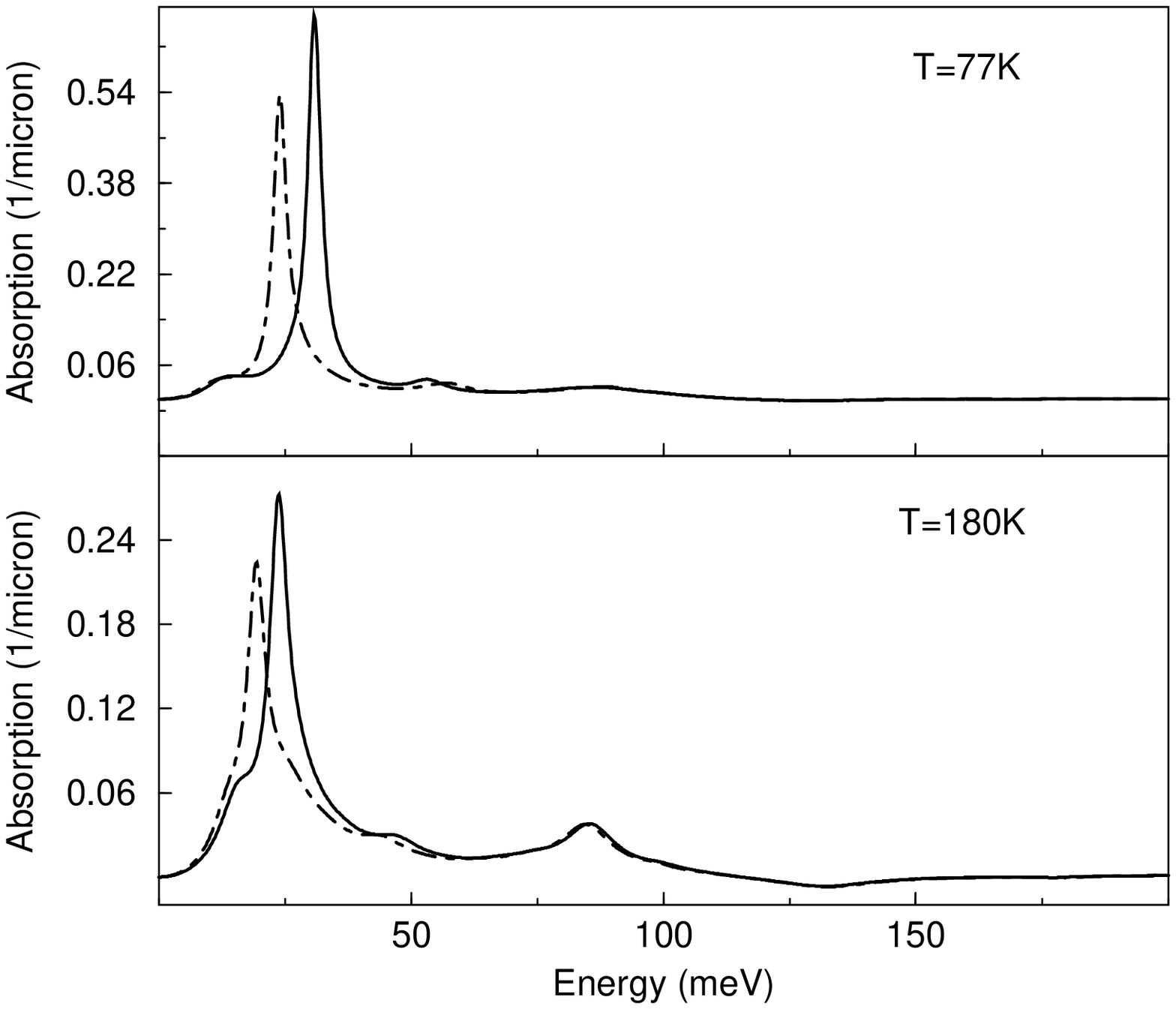}

\caption{Comparison of absorption spectra with (solid) and without 
(dot-dashed) Hartree-Fock Coulomb corrections for a bias
of 0.2 V/period .
The top and lower panels are respectively for
phonon bath temperatures T=77 and 180 K.} 
\label{figtemp}      
\end{figure}

We close our analysis with Fig.~\ref{figtemp} where we show the influence of temperature
on the calculated curves with (solid) and without (dot-dashed) many-body
effects. The top and lower panels are respectively for phonon bath
temperatures T=77 and 180 K. In both panels, the bias is V=0.2 V/period and
the carrier density N= 6.4 $\times$ $10^{11}$/cm$^{2}$. 

The rise in
temperature increases the spectral broadening and promotes a redistribution of
population difference, favouring the gain transitions  As a consequence, the
resulting  subband  shift is not so pronounced in the low energy side in 
comparison to T=77K.
However, since 
different transitions have different Coulomb corrections, the inclusion of
Coulomb effects at T=180 K affects the qualitative shape of the 
spectrum more strongly at higher temperatures. Structures that appear 
on the low energy side can not be seen with a free carrier model.

This manifests a further example of the
complicated interplay between the different levels in the
QCL structure, which is not present in simple two-level systems.

Another interesting remark is that, for larger dephasing, the counter-rotating
terms, usually very small and ignored
\cite{Li:03}, start to be relevant. The Coulomb
shifts have opposite directions for counter rotating terms than for
rotating wave contributions, and they should yield interesting effects
in structures specifically constructed to study their relevance.

\section{Summary}

In summary, our nonequilibrium many-body theory yields a surprising
contrast between the importance of Coulomb effects at the absorption
and gain region in quantum cascade structures.
While these effects are small in the gain region for the structure
considered,
strong features appear at THz absorption.

The many-body effects are influenced by both nonequilibrium occupation
functions and Coulomb matrix elements, which depend on wavefunction
overlap. Those two factors can be modified by means of an external bias
which ultimately allows us to control the overall strength of Coulomb
corrections in the structure.

\acknowledgments
The authors thank Science Foundation Ireland (SFI) and the Deutsche   
Forschungsgemeinschaft (DFG) for financial support of this work.  

\appendix
\section{Interpolation Formula for Coulomb Matrix Elements}

The Coulomb matrix elements in Eq.~\ref{Matrixelements} can be difficult to evaluate 
numerically, and we give here tractable expressions for the relevant terms. In
general, the Coulomb matrix elements are given by
\begin{widetext}   
\begin{eqnarray}
\label{ap1}      
V \left( { \mu \; \nu \; \lambda \; \beta \atop q } \right) &=&   
V^{2d}_{q} \; F \left( { \mu \; \nu \; \lambda \; \beta \atop q } \right), \\ \nonumber
F \left( { \mu \; \nu \; \lambda \; \beta \atop q } \right) &=&
\int dz \, dz'\, \phi^{\ast}_{\mu} (z)  \phi_{\nu} (z)    
\exp{ \left( - q | z - z' | \right) } \,   
\phi^{\ast}_{\lambda} (z') \phi_{\beta} (z'), \\ \nonumber
V^{2d}_{q} = \frac{2 \pi e^2} {\epsilon_0 A q}.
\end{eqnarray}  

\end{widetext}  
 
As pointed out in the text, $V^{\nu \mu \mu  \nu}_{0}$ matrix elements can be expressed exactly as a depolarisation length in the small wavenumber limit,

\begin{eqnarray}
L^{dep}_{\nu \mu} = \frac{ 9 \pi \epsilon_{0} A}{20 e^{2}} \, 
V^{\nu \mu \mu  \nu}_{0}.
\end{eqnarray}

The large wavenumber limit reads
\begin{eqnarray}
V \left( { \nu \; \mu \; \mu \; \nu \atop q \rightarrow \infty } \right) &=&  
V^{2d}_{q} \; \frac{1}{q \, b_{\mu \nu}},\\ \nonumber
\frac{1}{b_{\mu \nu}} &=& 2\int dz \, |\phi_{\mu} (z)|^{2} \, |\phi_{\nu} (z)|^{2}.    
\end{eqnarray}

Inspection of the two extremes, suggests an interpolation formula for the depolarisation-like
terms,
\begin{eqnarray}
V \left( { \nu \; \mu \; \mu \; \nu \atop q } \right) &=&  
V^{2d}_{q} \; \frac{q}{\frac{9 \pi^{2}}{10 L^{dep}_{\nu \mu}} + b_{\mu \nu} q^{2}}.
\end{eqnarray}

Similar considerations yield an analogous interpolation formula for the subband shift and
exchange matrix elements,
\begin{eqnarray}
V \left( { \nu \; \nu \; \nu \; \nu \atop q } \right) &=&  
V^{2d}_{q} \; \frac{1}{1 + \frac{L^{bg}_{\nu \nu}}{3} q}, \\ \nonumber
V \left( { \nu \; \nu \; \mu \; \mu \atop q } \right) &=&  
V^{2d}_{q} \; \frac{1}{1 + \frac{L^{ex}_{\nu \mu}}{2} q},
\end{eqnarray}

where
\begin{eqnarray}
L^{bg}_{\nu \nu} &=& 
\frac{3/2}{\int dz \, |\phi_{\nu} (z)|^{4}}, \\ \nonumber
L^{ex}_{\nu \mu} &=& 
\frac{1}{\int dz \, |\phi_{\mu} (z)|^{2} \, |\phi_{\nu} (z)|^{2}}.    
\end{eqnarray}

The lengths, $L^{dep}_{\nu \mu}$, $L^{ex}_{\nu \mu}$, $L^{bg}_{\nu \nu}$ have been defined so that they reduce to the
quantum well width if the wavefunctions used are the eigenstates of an infinite potential well.

\begin{table}     
\begin{tabular} {|l|l|l|l|l|r|r|r|r|r|}\hline 
Bias & ($\nu$,$\mu$) & $L^{dep}_{\nu \mu}$ & $\wp_{\nu \mu}$ & $n_{\nu}-n_{\mu}$ &  $E_{\mu}-E_{\nu}$ & $\Gamma_{\nu \mu}$ & $L^{ex}_{\nu \mu}$ & $L^{bg}_{\nu \nu}$ & $L^{bg}_{\mu \mu}$ \\ \hline\hline 
0.12 & (6,7)  & 25.8 & 4.99 & 2.44  & 13.3  & 5.3  & 14.9  & 25.0 & 30.3  \\ \hline
0.12 & (7,10) & 20.5 & 4.56 & 1.33  & 24.2  & 9.1  & 16.5  & 30.3 & 27.6 \\ \hline   
0.2  & (7,8)  & 19.3 & 4.23 & 3.19  & 23.9  & 3.1 & 17.2  & 20.4 & 27.4  \\ \hline 
0.2  & (8,10) & 18.5 & 3.96 & 0.87  & 25.3  & 9.0  & 14.0  & 27.4 & 31.6  \\ \hline 
0.2  & (4,9)  & 3.2  & 1.38 & -0.36 & 129.5 & 23.6 & 33.0  & 21.1 & 13.9 \\ \hline 
0.3  & (7,9)  & 12.3 & 2.50 & 2.08  & 32.8  & 6.2  & 12.8  & 18.5 & 22.2 \\ \hline 
0.3  & (8,11) & 11.7 & 2.92 & 1.07  & 73.9  & 14.4 & 12.2  & 14.2 & 33.7  \\ \hline 
0.3  & (6,7)  & 0.28 & 0.43 & -2.60 & 129.5 & 15.5 & 371.4 & 13.0 & 18.5 \\ \hline 
0.3  & (6,8)  & 4.5  & 1.68 & -1.08 & 139.2 & 19.3 & 20.6  & 13.0 & 14.2  \\ \hline  \end{tabular}     
\caption{Absorption/gain transition parameters. 
The first column is in   
Volts/period. ($\nu$,$\mu$) denotes the lower  and higher subbands in a transition,   
$n_{\nu}-n_{\mu}$ is the difference in subband population in units    
$10^{11}/\mathrm{cm}^{2}$.  $L^{dep}_{\nu \mu}$ 
 $L^{ex}_{\nu \mu}$ , $L^{bg}_{\nu \nu}$, and
$L^{bg}_{\mu \mu}$ are in nm and $\wp_{\nu \mu}$ in    
$e \times$nm. $E_{\mu}-E_{\nu}$ is the energy difference in meV at k=0 of   
Eq.~(4), i.e. including the static mean field corrections,   
but no further Coulomb effects. $\Gamma_{\nu \mu}$ = $\Gamma_{\nu} + \Gamma_{\mu}$ is the   
dephasing in meV.}   
\label{tab2}     
\end{table}

\begin{thebibliography}{15} 
\bibitem[\dag] PPresent Address: Department of Physics, 
University of Lund, Box 118, 22100 Lund, Sweden.     
\bibitem{FAI94a} J. Faist, F. Capasso, D.~L. Sivco, C. Sirtori, A.~L. Hutchinson, and A.~Y. Cho ,Science {\bf 264},  553  (1994).
     
      
\bibitem{KOE02} R. Kohler, A. Tredicucci, F. Beltram, H. E. Beere, E. H. Linfield, A. G. Davies, D. A. Ritchie, R. C. Iotti and F. Rossi, Nature, {\bf417}, 156 (2002).

  
\bibitem{Zim} R. Zimmermann, {\em Many Particle Theory of Highly Excited   
Semiconductors} (Teubner, Leipzig, 1987).   
      
\bibitem{Helm:99} M. Helm, in {\em Intersubband Transitions in Quantum Wells:  Physics and Device Applications}, edited by E.~R. Weber and R.~K. Willardson   
(Academic Press), {\bf 62}, 1 (1999).   


\bibitem{Li:03} J.~Li and C.~-Z.~Ning, Phys. Rev. Lett. {\bf 91}, 
097401 (2003).  

 
\bibitem{Ines:03} I.~Waldm\"uller, M. Woerner, Jens F\"orstner, and A. Knorr 
Phys. stat. sol. {\bf b 238}, 474 (2003).

 
\bibitem{NIKO97} D. E. Nikonov, A. Imamoglu, L. V. Butov, and 
H. Schmidt,
Phys. Rev. Lett. {\bf 79} 4633 (1997).   


\bibitem{Ning:02} J. Li, C.~Z.~Ning, D.~C.~Larraber, G.~A.~Khodaparast, J.~Kono, K.~Ueda, Y.~Nakajima, S.~Sasa, and M.~Inoue, in Progress in Semiconductors   
II edited by B.D. Weaver, (Proceedings of the MRS) {bf 744}, 571 (2003).

\bibitem{Gumbs:95} G. Gumbs, D. Huang, and J.P. Loehr,
 Phys.~Rev.~B{\bf 51}, 4321 (1995).

\bibitem{Faleev:02} S.V. Faleev, and M.I. Stockman, Phys.~Rev.~B {\bf   
66}, 085318 (2002).    

\bibitem{PER98} M.~F. {Pereira, Jr.} and K. Henneberger, Phys.~Rev.~B {\bf   
58}, 2064 (1998). 

\bibitem{Tsu:00} S.~Tsujino, M.~R\"ufenacht, H.~Nakajima, T.~Noda,C.~Metzner, and H.~Sakaki, Phys.~Rev.~B {\bf 62}, 1560 (2000).  

\bibitem{Luin:01} S.~Luin, V.~Pellegrini, F.~Beltram, X.~Marcadet, and C.~Sirtori, Phys.~Rev.~B {\bf 64}, 041306 (2001).

\bibitem{Liu:00} H. C. Liu and A. J. Spring Thorpe, Phys.~Rev.~B {\bf 61}, 15629 (2000).
   
\bibitem{SIR98}C.~Sirtori, P.~Kruck, S.~Barbieri, P.~Collot, J.~Nagle,
M.~Beck, J.~Faist, and U.~Oesterle, Appl.~Phys.~Lett. {\bf 73}, 3486
  (1998).
\bibitem{GIE03} M. Giehler, R.~Hey, H.~Kostial, S.~Cronenberg, L.~Schrottke, and H.~T.~Grahn, Appl.~Phys.~Lett. {\bf 82},
  671 (2003).
\bibitem{LEE02a} S. C. Lee and A. Wacker, Phys.~Rev.~B {\bf 66}, 245314 (2002).   
\bibitem{COM02} F.~Compagnone, A.~Di~Carlo, and P.~Lugli, Appl.~Phys.~Lett. {\bf 80}, 920 (2002).   
\bibitem{CHU92} S.~L.~Chuang, M.~S.~C.~Luo, S.~Schmitt-Rink and A.~Pinczuk,
Phys.~Rev.~B {\bf 46}, 1897 (1992).   
\bibitem{IOT01a} R.~C.~Iotti and F.~Rossi, Phys.~Rev.~Lett. {\bf 87}, 146603   
(2001).   
\end{thebibliography}
 \end{document}